\documentclass[12pt,preprint]{aastex}

\def\hkpc{$~h_{70}^{-1}$ kpc}

\shorttitle{The Origin of the Mass-Metallicity Relation}
\shortauthors{Ellison et al.}

\begin{document}


\title{Clues to the Origin of the Mass-Metallicity Relation: Dependence
on Star Formation Rate and Galaxy Size.}

\author{Sara L. Ellison\altaffilmark{1},
David R. Patton\altaffilmark{2,3},
Luc Simard\altaffilmark{4},
\& 
Alan W. McConnachie\altaffilmark{1}
}

\altaffiltext{1}{Dept. of Physics \& Astronomy, University of Victoria,  
3800 Finnerty Rd, Victoria, V8P 1A1, British Columbia, Canada,
sarae@uvic.ca, alan@uvic.ca}

\altaffiltext{2}{Department of Physics \& Astronomy, Trent University, 
1600 West Bank Drive, Peterborough, Ontario, K9J 7B8, Canada, dpatton@trentu.ca}

\altaffiltext{3}{Visiting Researcher, Dept. of Physics \& Astronomy, 
University of Victoria,  
3800 Finnerty Rd, Victoria, V8P 1A1, British Columbia, Canada}

\altaffiltext{4}{National Research Council of Canada,
Herzberg Institute of Astrophysics, 5071 West
Saanich Road, Victoria, British Columbia, V9E 2E7, Canada,
luc.simard@nrc.ca}




\begin{abstract}
We use a sample of 43,690 galaxies selected from the Sloan Digital Sky
Survey Data Release 4 to study the systematic effects of specific star
formation rate (SSFR) and galaxy size (as measured by the half light
radius, $r_h$) on the mass-metallicity relation.  We find that
galaxies with high SSFR or large $r_h$ for their stellar mass have
systematically lower gas phase-metallicities (by up to 0.2 dex) than galaxies
with low SSFR or small $r_h$.  We discuss possible origins for these
dependencies, including galactic winds/outflows, abundance gradients,
environment and star formation rate efficiencies.
\end{abstract}




\section{Introduction}

The well-established luminosity-metallicity relation (LZR) that exists over
at least 10 magnitudes in M$_B$ (e.g., Skillman, Kennicutt \& Hodge 1989;
Zaritsky, Kennicutt \& Huchra 1994; Salzer et al. 2005; Lee et al. 2006)
has been recently confirmed to be a manifestation of a more fundamental
stellar mass-metallicity relation (MZR; Lequeux et al. 1979;
Tremonti et al. 2004).    Although there
is evidence that galaxy interactions may affect the 
normalization of the LZR and MZR (Kewley et al. 2006; Ellison et al. 
2007; Rupke et al. 2007), the MZR appears to be independent of
large-scale environment (Mouhcine, Baldry \& Bamford 2007) and
remains intact out to $z \sim 2.5$ (Savaglio et al. 2005; Erb et al. 2006).

Several origins of this apparently fundamental relation 
have been proposed.  Many of these models
invoke winds as the fundamental driver of the MZR (e.g., Kobayshi,
Springel \& White 2007) and observational evidence indicates
that gas ejection and/or accretion could be important (Tremonti
et al. 2004; Gallazzi et al. 2005).
However, the effects of supernova feedback on star formation
efficiency and metal-poor gas infall have also been cited as important
factors (Brooks et al. 2007; Finalator \& Dav\'e 2007).  
In this paper, we investigate the dependence of the MZR on various
physical parameters that may shed light on the underlying mechanism
that shapes it.  

We adopt a concordance cosmology of $\Omega_{\Lambda} = 0.7$, $\Omega_M = 0.3$,
$H_0 = 70$ km/s/Mpc where applicable.

\section{Data Sample}

We use the Sloan Digital Sky Survey Data Release 4
(SDSS DR4) to compile a sample of star-forming
galaxies with spectra suitable for metallicity determinations.  
The galaxy sample is very
similar to the `control' sample used in the study of close galaxy
pairs by Ellison et al. (2007) where a full description of the
sample selection is presented.  In brief, the sample consists
of galaxies with extinction corrected Petrosian magnitudes in the 
range 14.5 $< r$ $\le$ 17.77 with strong emission lines. 
Galaxies must be classified as star-forming and not AGN dominated, 
according to the line diagnostic criteria given in Kewley et al. 
(2001)\footnote{Using the AGN classification scheme of Kauffmann
et al. (2003a) yields identical science results, but results
in a sample of only 38703 galaxies, a reduction of $\sim$ 10\%.
This is in agreement with Kewley \& Ellison (2007) who find
that the AGN removal method does not affect the shape of the MZR.}.
The galaxies must also have available metallicities from the
Kewley \& Dopita (2002) `recommended' method, masses determined
from spectral synthesis modelling (Kauffmann et al. 2003b),
aperture corrected 
SFRs from Brinchmann et al. (2004) and photometric/morphological
parameters such as $r$-band half light radii ($r_h$) and bulge-to-total
fractions (B/T) from Simard (in preparation) derived from GIM2D
(see Simard et al. 2002 for details on the fitting procedure).  
The GIM2D $r_h$ values are in excellent agreement with the
values derived from Sersic fits (Blanton et al. 2005).
In contrast to Ellison et al. (2007), we do not impose an upper redshift
limit and we require the fiber covering fraction (CF: the ratio of the
g band Petrosian to fiber fluxes) to
be CF $\ge$ 20\% in order to minimize the aperture effects on
metallicity (e.g., Kewley et al. 2005; Ellison \& Kewley 2005; 
Kewley \& Ellison 2007).  The final sample consists of 43,690 galaxies
with a median redshift of $z = 0.086$.

\section{Results}

\begin{figure}[t]
\centerline{\rotatebox{270}{\resizebox{6cm}{!}
{\includegraphics{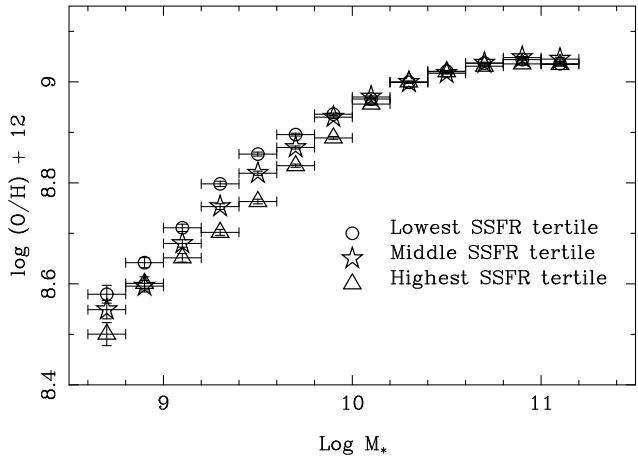}}}}
\caption{\label{ssfr} The binned mass-metallicity relation for $\sim$ 44,000
SDSS galaxies from the DR4, divided by specific star formation rate, 
log SSFR = log [SFR (M$_{\odot}$yr$^{-1}$)/M$_{\star}$(M$_{\odot}$)],
tertiles in each mass bin.  Open circles, stars and triangles represent 
the lowest, intermediate and highest SSFRs in units of log (yr$^{-1}$)
respectively. The divisions
between the SSFR tertiles are log SSFR $< -9.8$ yr$^{-1}$, $-9.8 <$ SSFR 
$< -9.6$
yr$^{-1}$ and SSFR $> -9.6$ yr$^{-1}$ for the mass bin centered at log
M$_{\star}$ = 9.5 M$_{\odot}$ and SSFR $< -10.0$ yr$^{-1}$, $-10.0 <$ SSFR $< -9.7$
yr$^{-1}$ and SSFR $> -9.7$ yr$^{-1}$ for the mass bin centered at log
M$_{\star}$ = 10.5 M$_{\odot}$.  The vertical error bars are the standard
error on the mean.}
\end{figure}

In Figure \ref{ssfr} we show the MZR for our sample of $\sim$ 44,000
SDSS galaxies divided by specific star formation rate (SSFR).  
For reference, the median H$\alpha$ + [NII] equivalent width (EW),
which is a measure of the present to past-average star formation
rate (see Figure 3 of Kennicutt et al. 1994),
of galaxies with log SSFR $> -9.5$ is 56 \AA.  
Since SSFR itself depends on mass, with lower values at higher masses, 
we plot the SSFRs by tertiles
calculated separately in each mass bin.    At high stellar masses (log
M$_{\star} >10$ M$_{\odot}$) the MZR exhibits no dependence on
SSFR.   At lower stellar masses, there is a tendency for galaxies with
higher SSFR to have lower metallicities for a given stellar mass.  The
offset in metallicity from the highest to lowest SSFR bins is
$\sim 0.10$ -- 0.15 dex in the stellar mass range $9 <$ log M$_{\star}
<$ 10 M$_{\odot}$ and is significant at greater than 5$\sigma$ (the
standard error on the mean) in these mass bins.

\begin{figure}[t]
\centerline{\rotatebox{270}{\resizebox{6cm}{!}
{\includegraphics{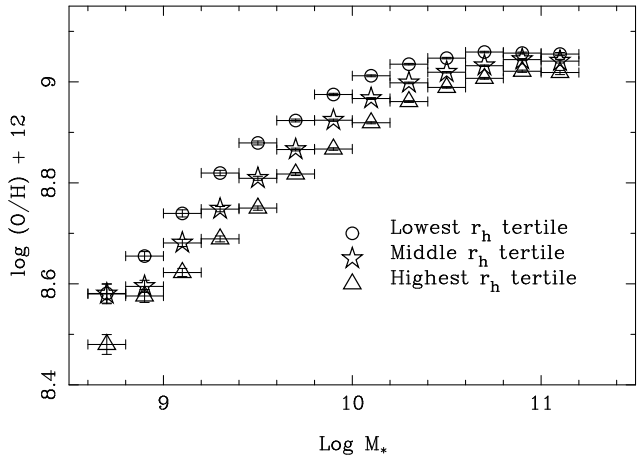}}}}
\caption{\label{rh} The binned mass-metallicity relation for $\sim$ 44,000
SDSS galaxies from the DR4, divided by half light radius ($r_h$) tertiles
in each mass bin.  Open circles, stars and triangles represent 
the lowest, intermediate and highest $r_h$ respectively. The divisions
between the $r_h$ tertiles are $r_h < 2.2$ \hkpc, $2.2 < r_h < 3.3$
\hkpc\ and $r_h > 3.3$ \hkpc\ for the mass bin centered at log M$_{\star}$
= 9.5 M$_{\odot}$ and $r_h < 4.0$ \hkpc, $4.0 < r_h < 5.5$
\hkpc\ and $r_h > 5.5$ \hkpc\ for the mass bin centered at log M$_{\star}$
= 10.5 M$_{\odot}$.  The vertical error bars are the standard error on the mean.}
\end{figure}

\begin{figure}[t]
\centerline{\rotatebox{270}{\resizebox{6cm}{!}
{\includegraphics{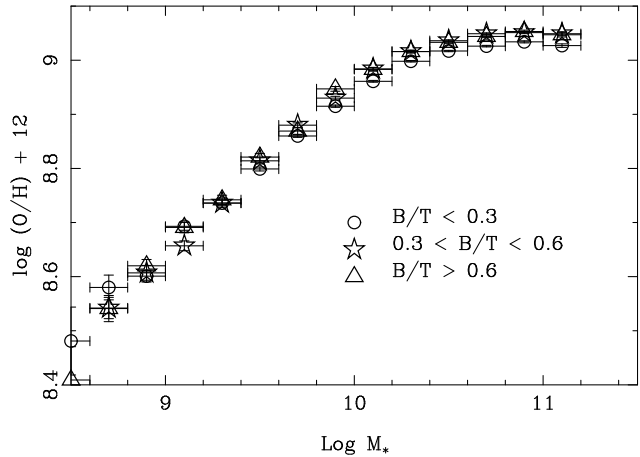}}}}
\caption{\label{bt} The binned mass-metallicity relation for $\sim$ 44,000
SDSS galaxies from the DR4, divided by $r$-band bulge-to-total fraction
(B/T).  Open circles, stars and triangles represent B/T$<0.3$,
$0.3 \le$ B/T $<0.6$ and B/T $\ge 0.6$ respectively.  The vertical 
error bars are the standard error on the mean.}
\end{figure}

In Figure \ref{rh} we again show the MZR for our sample of 
SDSS galaxies, but now divided by half light radius.  
Since $r_h$ itself depends on mass, with a tendency towards
higher values at higher masses,
we plot the $r_h$ by tertiles calculated separately in each mass bin.
There is an offset in metallicity between the galaxies with the 
smallest optical
extents and those with the largest half light radii, with a
range in the binned values of between 0.05 -- 0.2 dex.  
A similar result was noted by Tremonti et al. (2004), who
found that, for a fixed M$_{\star}$, galaxies with higher mass surface
density (i.e. smaller $r_h$)  had higher 
metallicities, equivalent to the trend in Figure \ref{rh}.
The significance of the metallicity offset between $r_h$ tertiles
in the range $9 <$ log M$_{\star} <$ 10 M$_{\odot}$ is greater than 10$\sigma$ 
in a given mass bin.    However, despite the offsets shown in Figures 
\ref{ssfr} and \ref{rh}, the dependence of the MZR on SSFR and $r_h$ 
is not a significant cause of scatter in the relation as a whole
(i.e., when all galaxies are included).    
The decrease in scatter for the MZR binned by $r_h$
compared to the full sample is less than 10\%, indicating that the
scatter is dominated by either an additional parameter, observational
errors or the accuracy of the metallicity calibration.  As was
the case for the offsets in SSFR, in the stellar mass bins
with sufficient statistics, we see a tendency for larger
offsets in the MZR between $r_h$ bins at lower stellar masses.
The offset in the MZR when binned by $r_h$ is not due to a
correlation between size and morphology.  Figure \ref{bt} shows
that the MZR is not significantly different for galaxies with
different bulge fractions.

Trends in the MZR with size have previously been reported in
the literature, e.g., Hoopes et al. (2007) for local UV luminous
galaxies and Ellison et al. (2007) for galaxies with close companions.  
However,
both of these studies have found that more compact galaxies
have lower metallicities for their mass, compared with more extended
galaxies.  This trend is opposite to what we see in 
Figure \ref{rh}.  One explanation is that the MZR offsets seen
by Hoopes et al. (2007) and Ellison et al. (2007) are the result of
merger induced activity, as supported by the MZR offsets observed
in luminous and ultra-luminous infra-red galaxies
by Rupke et al. (2007).  We discuss alternative origins for the 
trends in Figures \ref{ssfr} and \ref{rh} in the next section.

Before discussing the possible physical origin of the MZR's dependence
on SSFR and $r_h$, it is interesting to consider the practical impact
of these dependences on comparisons between local and high redshift
versions of the MZR.  Once aperture effects and the use of a single
metallicity diagnostic (e.g., Ellison \& Kewley 2005; Kewley et al.
2005; Savaglio et al. 2005; Erb et al. 2006; Kewley \& Ellison 2007)
are accounted for, there is an offset to lower
metallicities by about 0.3 dex for a galaxy of given mass at $z \sim$
2.5 compared with $z = 0$ (Erb et al. 2006).  However, our results
indicate that the observational selection biases of Lyman break galaxy 
(LBG) samples may also play a role.  Due to the
(1+$z)^4$ surface brightness dimming effect, it is easier to detect
and study high redshift galaxies when they are compact.  The typical
$r_h$ of a high redshift LBG is $\sim$ 2 \hkpc\
(Dickinson 2000), which could potentially result in an upward shift in
their observed MZR (see Figure \ref{rh}).    
Conversely, LBGs tend to have quite
high SSFRs; all of the binned SSFRs in Erb et al. (2006) 
have values of log SSFR $\ge -9.3$ yr$^{-1}$, which would result
in a downwards shift in their MZR (see Figure \ref{ssfr}) relative
to a more complete local sample.  
To assess the combination of these opposite trends, we 
compare the MZR of our full sample of 43,690 galaxies with 
that of the 473 galaxies with $r_h \le$ 2 \hkpc\ and SSFR $\ge$ 9.3
yr$^{-1}$ (the `LBG-like' sample).   
We find that the LBG-like sample has a small negative offset, i.e., to
lower metallicities, by $\sim$ 0.05 dex at stellar masses log
$M_{\star} <$ 10 M$_{\odot}$; the effect of SSFRs therefore appears to
dominate in this regime.  For stellar masses
$M_{\star} > $10.2 M$_{\odot}$, where the MZR is independent of SSFR
(see Figure \ref{ssfr}), the effect of the $r_h$ dependence is dominant
and the MZR of the LBG-like galaxies is positively offset by $\sim$0.05
dex, as expected from Figure \ref{rh}.  However, these shifts are
much smaller than the observational scatter in either the
low or high redshift MZR.

\section{Discussion: Possible Interpretations of Mass-Metallicity Trends}

\textbf{Abundance gradients.}
Radial gradients in galactic oxygen abundance are well established
and have typical magnitudes of $-$0.01 to $-$0.05 dex kpc$^{-1}$ (i.e.,
higher metallicities in the centers of galaxies), with $-$0.03
dex kpc$^{-1}$ typical for local spiral galaxies (e.g., Zaritsky
et al. 1994).  A combination
of fiber covering fraction and abundance gradient effects could therefore
cause an apparent dependence of the MZR on $r_h$.  The median CFs 
are 0.40, 0.29 and 0.24 for the smallest, intermediate and largest
$r_h$ tertiles respectively at log M$_{\star} = 10$ M$_{\odot}$.  
For a fixed abundance gradient, galaxies with higher fiber covering 
fractions are expected to have lower observed metallicites, since the 
fiber probes more of the outer disk.  Hence, an aperture bias would
lead to lower metallicities in small $r_h$ galaxies (whose CF is
large), opposite to what we find in Figure \ref{rh}.
However, if the steepness of abundance gradients depends on $r_h$
\textit{for a given mass}, then aperture effects may play a role.
The dependence of radial abundance gradients is complex and may
depend on morphology (such as the presence of a bar, e.g.,
Martin \& Roy 1995), as well as size and galactocentric radius
(Magrini et al. 2007).  It is also not clear how the magnitude of an abundance
gradient varies with $r_h$.  On the one hand, nearby dwarfs tend to
have very flat abundance gradients (e.g., Lee, Skillman \& Venn 2006).
Conversely, Prantzos \& Boissier (2000) have shown that spiral galaxies
with smaller disks may have steeper gradients than those with 
large disks.  

Nonetheless, we can still consider whether the
typical differences in abundance gradients between galaxies of different
sizes could yield offsets in the
MZR commensurate with the observed segregation in Figure \ref{rh}.
At the median redshift of our sample,
the SDSS fiber radius corresponds to $\sim$ 2 \hkpc\ and the
typical abundance offset between large and small $r_h$ galaxies
in Figure \ref{rh} is 0.1--0.2 dex.  Since the difference between
the abundance gradients of large disks and either small disks
or dwarfs is small, typically $\ll$ 0.05 dex kpc$^{-1}$, these 
differences seem unlikely to be the cause of the offsets in 
Figure \ref{rh}.  We can also test empirically for the effect of 
varying abundance gradients by looking for segregation in the MZR based on
both cuts in $r_h$ and covering fraction, where the presence of
$r_h$-dependent gradients plus aperture effects
would manifest themselves as a smaller segregation by $r_h$ as
the covering fraction increased.  We find no evidence for such an
effect (although the median CF of our sample is only 30\%).
We conclude that abundance gradients are unlikely to explain
(all) of the $r_h$ dependence seen in Figure \ref{rh}.

\textbf{Galactic winds.}
Winds may seem like a natural explanation for the dependence
of the MZR on $r_h$ and SSFR.  For example, the results shown in 
Figure \ref{ssfr} could be explained by a wind model in which
higher SSFRs lead to more efficient evacuation of metals and hence 
a downward shift in the MZR.  However, this explanation requires
an assumption of instantaneous mixing of metals into the cool interstellar
medium (ISM).  In reality,
the metals produced in the current episodes of star formation 
(i.e., those measured by the SSFR) have not yet sufficiently
cooled to be traced by the HII phase.
Moreover, whilst galaxies with smaller $r_h$ 
for a given M$_{\star}$ will have more centrally-concentrated 
stellar masses and higher surface gravities, in the models od
Finlator \& Dav\'e (2007) the wind velocities 
are always easily high enough to
escape the potential wells of the galaxies' gravity.  If true in
practice, this would indicate that gravitational potential alone is
not sufficient to inhibit metal loss through winds in more centrally
concentrated galaxies.  An alternative is that the hot wind 
entrains more of the cool ambient ISM.  However,
Dalcanton (2007) has used an analytic model of gas infall/outflow
to show that even a modest amount of star formation following
metal-loss through winds quickly erases the signature of
such an event.

\textbf{Infall of metal-poor gas.}
In the model of Finlator \& Dav\'e (2007), it is proposed 
that all galaxies have 
a fundamental equilibrium metallicity ($Z_{\rm eq}$)
for a given mass.  Deviations from this value are caused
by the inflow of pristine gas from the IGM, but $Z_{\rm eq}$ is 
eventually restored by the effects of star formation, winds and mass loss.
In this scenario, the low
metallicities seen in Figure \ref{ssfr} could be explained by recent
inflow of metal poor gas which shifts the MZR from its equilibrium
position.  In response to the deposition of fresh fuel, which in
turn increases the gas surface density, the galaxy will experience
an increase in its SFR.  Indeed, the same qualitative dependence of
the MZR on SSFR is seen in the models (Finlator, private communication).
The dependence on SSFR is more pronounced at lower
masses since it is more difficult to perturb a high mass galaxy from
its equilibrium position.  

On the other hand, the timescale arguments of the infall model
appear inconsistent with Figure \ref{rh}.
The time taken to recover from the injection of metal
poor gas, and return to the equilibrium metallicity, is the dilution time,
$t_d = M_{\rm gas}/\dot{M}_{\rm acc}$, where $M_{\rm gas}$ 
is the gas mass and
$\dot{M}_{\rm acc}$ is the gas accretion rate.  If $t_d < t_{\rm dyn}$ 
(where $t_{\rm dyn}$ is the dynamical time)
then the galaxy `recovers' its equilibrium
metallicity promptly, leading to very little scatter in the MZ relation.  
Conversely, if $t_d > t_{\rm dyn}$, then the galaxy struggles to 
recover promptly from inflows, leading to a large scatter in the MZR,
due to galaxies which are displaced to lower metallicities.  
Since $t_{\rm dyn} \propto \sqrt{r^3/M_{\rm gal}}$, then, all
other things being equal,
$t_d/t_{\rm dyn} \propto r^{-3/2}$.  Therefore, galaxies with 
smaller radius may be expected to have larger $t_d/t_{\rm dyn}$,
hence taking longer to recover their equilibrium metallicity after
an injection of metal poor gas.  This effect should manifest itself
in Figure \ref{rh} as a shift towards lower metallicities for 
smaller $r_h$ galaxies if a large fraction of them have experienced 
recent infall. Indeed, such an effect is seen in the galaxy pairs
sample of Ellison et al. (2007).  However, the inverse of the
expected $r_h$ dependence is seen in the galaxy sample considered
here, with
small $r_h$ galaxies exhibiting the highest metallicities for a given mass.

\textbf{Environmental dependence.}  This also seems an unlikely explanation
for the trends in Figures \ref{ssfr} and \ref{rh}.  
Mouhcine et al. (2007) have recently shown that the MZR 
does not depend sensitively on large-scale environment.  Whilst
galaxy interactions can perturb the LZR and MZR 
(Kewley et al. 2006; Rupke et al. 2007), the dependence of
this effect on $r_h$ is the opposite to that seen in Figure
\ref{rh} (Ellison et al. 2007).  Moreover, Ellison et al. (2007)
attributed the $r_h$ dependence of the MZR in galaxy pairs
to the stabilizing effect of a bulge (e.g., Mihos \& Hernquist 1994, 
1996; Cox et al. 2007).  We
find that there is no significant difference in the MZR 
for galaxies with different bulge fractions, see Figure \ref{bt}.  
In turn, this indicates that although early-type galaxies
contain a higher fraction of the low $z$ metals budget (e.g., Gallazzi
et al. 2007), morphology does not appear to segregate the MZR.

\textbf{Star formation efficiencies.}
Differing star formation efficiencies (SFE) is another possible 
explanation for the trends in Figures \ref{ssfr} and 
\ref{rh}.   If the gas (as well as light) is more centrally
located in small $r_h$ galaxies, and has a higher surface density,
this could lead to higher SFE in the past.  In turn, this would
yield higher present-day gas-phase metallicities but lower current SSFRs
if more of the gas has been depleted as a result of the past star
formation.  A prediction of this scenario, which explains the 
trends in Figures \ref{ssfr} and \ref{rh}, is that the galaxies
with the highest metallicities at a given mass should also be the reddest
and have the lowest H$\alpha$+[NII] EW (a measure of present to past-average 
star formation, Kennicutt et al. 1994).  
These effects are indeed present in the data. Based on the GIM2D
bulge and disk magnitudes, we find that galaxies in the 
reddest $(g-r)_{disk}$ tertile at log M$_{\star}$ = 9.5 M$_{\odot}$
have a median metallicity 0.2
dex greater than the bluest tertile of galaxies in the same mass
bin.  There is no segregation in the MZR for bulge colors. 
The offset in metallicity between the highest and lowest H$\alpha$+[NII]
EW tertiles at the same stellar mass is 0.1 dex.
Kauffmann et al. (2003b) have similarly concluded that star formation
histories are most sensitive to surface mass density.

\section{Conclusions}

We have investigated how the galactic MZR depends on observable parameters 
such as the half light radius, SSFR and morphology and discuss various
scenarios to explain the dependencies.  
We find that, at a given stellar mass, the MZR is offset towards 
lower metallicities (by up to 0.2 dex)
for galaxies with larger half light radii and
higher specific star formation rates, but that there is no
significant dependence on bulge fraction.  These
dependencies have little impact on the overall scatter in the MZR
and the basic shape of the relation exists for all subsets of $r_h$
and SSFR.  We conclude that environment and abundance gradients
are unlikely to account for (all) of the $r_h$ and SSFR dependence
of the MZR.  Infall of metal-poor gas or metal-enriched outflows
also seem unlikely explanations based on timescale arguments.  Of
the possibilities considered here, a sensitivity to star formation
efficiency is the most plausible reason for the dependence of
the MZR on $r_h$ and SSFR.

\acknowledgements

We are grateful to the Munich group for making their SDSS galaxy
catalogs publicly available and to Lisa Kewley for providing her metallicity
calibrations.  We have benefitted from insightful discussions
with  Alyson Brooks, Romeel Dav\'e, Kristian Finlator and Fabio Governato
and from extremely useful feedback from the referee.  SLE and DRP are 
supported by NSERC Discovery Grants.


\begin{thebibliography}{}

\bibitem[Blanton et al. (2005)]{bla05}
        Blanton, M. R., et al. 2005, AJ, 129, 2562

\bibitem[Brinchmann et al. 2004]{bri04}
        Brinchmann, J., Charlot, S., White, S. D. M., Tremonti, C., 
	Kauffmann, G., Heckman, T., Brinkmann, J.,2004, MNRAS, 351, 1151  

\bibitem[Brooks et al. (2007)]{bro07}
        Brooks, A. M., Governato, F., Booth, C. M., Willman, B., 
	Gardner, J. P., Wadsley, J., Stinson, G., Quinn, T.,
	 2007, ApJ, 655, L17

\bibitem[Cox et al. (2007)]{cox07}
        Cox, T. J., Jonsson, P., Somerville, R. S., Primack, J. R.,
	Dekel, A., MNRAS, 2007, submitted

\bibitem[Dalcanton (2007)]{dal07}
        Dalcanton, J. J., 2007, ApJ, 658, 941

\bibitem[Dickinson (2000)]{dic00}
         Dickinson, M., 2000, Philos. Trans. R. Soc. London A, 
	 358, 2001 

\bibitem[Ellison et al. (2007)]{sdss2}
        Ellison, S. L., Patton, D. R., Simard, L., 
	McConnachie, A. W., 2007, AJ, submitted

\bibitem[Ellison \& Kewley (2005)]{ek05}
        Ellison, S. L., Kewley, L. J.,	2005, pg 53,
	Proceedings of "The Fabulous Destiny of Galaxies; 
	Bridging the Past and Present", Eds Le Brun, Mazure, 
	Arnouts, Burgarella

\bibitem[Erb et al. (2006)]{erb06}
        Erb, D. K., Shapley, A. E., Pettini, M., Steidel, C. C., 
	Reddy, N. A., Adelberger, K. L.,  2006, ApJ, 644, 813 

\bibitem[Finlator \& Dave (2007)]{fd07}
        Finlator, K., \& Dav\'e, R., 2007, MNRAS, submitted, arXiv:0704.3100

\bibitem[Gallazzi et al. (2007)]{gal07}
        Gallazzi, A., Brinchmann, J., Charlot, S., White, S. D. M.,
	2007, MNRAS, submitted, arXiv:0708.0533

\bibitem[Gallazzi et al. (2005)]{gal05}
        Gallazzi, A.,  Charlot, S., Brinchmann, J., White, S. D. M.,
	Tremonti, C. A., 2005, MNRAS, 362, 41

\bibitem[Hoopes et al. (2007)]{hoo07}
        Hoopes, C. G., et al., 2007, ApJS, in press, arXiv:astro-ph/0609415v1

\bibitem[Kauffmann et al. (2003a)]{kau0a3} 
	Kauffmann, G., et al., 2003a, MNRAS, 346, 1055

\bibitem[Kauffmann et al. (2003b)]{kau03b} 
	Kauffmann, G., et al., 2003b, MNRAS, 341, 33

\bibitem[Kennicutt et al. (1994)]{ken94}
        Kennicutt, R. C., Tamblyn, P., Congdon, C. E., 1994
	ApJ, 435, 22

\bibitem[Kewley \& Dopita (2002)]{kd02}
        Kewley, L. J., \& Dopita, M. A., 2002, ApJS, 142, 35

\bibitem[Kewley \& Ellison (2007)]{ke07}
        Kewley, L. J., \& Ellison, S. L., 2007, ApJ, submitted

\bibitem[Kewley et al (2001)]{kew01}
        Kewley, L. J., Heisler, C. A., Dopita, M. A., Lumsden, S.,
	2001, ApJS, 132, 37

\bibitem[Kewley et al. (2005)]{kjg05}
         Kewley, L. J., Jansen, R. A., Geller, M. J., 
	  2005, PASP, 117, 227

\bibitem[Kobayashi, Springel \& White (2007)]{ksw07}
        Kobayashi, C., Springel, V., \& White, S. D. M., 
	2007, MNRAS, 376, 1465

\bibitem[Lee et al. (2006)]{lee06}
        Lee, H., Skillman, E. D., Cannon, J. M., Jackson, D. C., 
	Gehrz, R. D., Polomski, E. F., Woodward, C. E.,
	 2006, ApJ, 647, 970

\bibitem[Lee, Skillman \& Venn (2006)]{lsv06}
        Lee, H., Skillman, E. D., Venn, K. A., 2006, ApJ, 642, 813

\bibitem[Lequeux et al. (1979)]{leq79}
         Lequeux, J., Peimbert, M., Rayo J. F., Serrano, A., 
	 Torres-Peimbert, S., 1979, A\&A, 80, 155

\bibitem[Magrini et al. (2007)]{mag07}
        Magrini, L., Vilchez, J. M., Mampaso, A., Corradi, R. L. M., 
	Leisy, P.,  2007, A\&A, 470, 865

\bibitem[Martin \& Roy (1995)]{mr95}
        Martin, P., \& Roy, J.-R., 1995, ApJ, 445, 161

\bibitem[Mihos \& Hernquist (1994)]{mh94}
        Mihos, C., \& Hernquist, L., 1994, ApJ, 425, L13

\bibitem[Mihos \& Hernquist (1996)]{mh96}
        Mihos, C., \& Hernquist, L., 1996, ApJ, 464, 641

\bibitem[Mouhcine, Baldry \& Bamford (2007)]{mbb07}
        Mouhcine, M., Baldry, I. K., Bamford, S. P., 2007, MNRAS
	accepted, arXiv:0709.3794

\bibitem[Prantzos and Boissier 2000]{pb00}
	Prantzos, N., Boissier, S., 2000, MNRAS, 313, 338

\bibitem[Rupke, Veilleux \& Baker (2007)]{rvb07}
         Rupke D. S. N., Veilleux, S., \& Baker, A. J., 2007, 
	 ApJ, accepted, arXiv:0708.1766 

\bibitem[Salzer et al. (2005)]{sal05}
         Salzer, J. J., Lee, J. C., Melbourne, J., Hinz, J. L., 
	 Alonso-Herrero, A., Jangren, A., 2005, ApJ, 624, 661

\bibitem[Savaglio et al. (2005)]{sav05}
        Savaglio, S., et al.,  2005, ApJ, 635, 260 

\bibitem[Simard et al. (2002)]{sim02}
        Simard, L., Willmer, C. N. A., Vogt, N. P., Sarajedini, V. L., 
	Phillips, A. C., Weiner, B. J., Koo, D. C., Im, M., 
	Illingworth, G. D., Faber, S. M., 2002, ApJS, 142, 1

\bibitem[Skillman, Kennicutt \& Hodge (1989)]{skh89}
        Skillman, E. D., Kennicutt, R. C., \& Hodge, P. W., ApJ, 1989, 
	347, 875 

\bibitem[Zaritsky, Kennicutt \& Huchra (1994)]{zkh94}
        Zaritsky, D., Kennicutt, R. C., Jr., Huchra, J. P., 1994,
	ApJ, 420, 87

\end{thebibliography}
\end{document}